\documentclass[10pt,twocolumn,a4paper,titlepage]{article}
\usepackage{latex8}
\usepackage{times}
\usepackage{epsfig}
\usepackage{psfrag}
\usepackage{psfig}
\usepackage{amsmath}
\usepackage{latexsym}
\usepackage[T1]{fontenc}
\usepackage[english]{babel}
\newcommand{\remove}[1]{}

\newtheorem{theorem}{\bf Theorem}
\newtheorem{lemma}{\bf Lemma}

\newtheorem{definition}{\bf Definition}

\newenvironment{proof}{\noindent \bf Proof:\rm}{\hspace*{\fill}$\Box$\vspace{1ex}}

\begin{document}

\onecolumn

\begin{titlepage}
\title{
  \raisebox{30mm}[0mm][0mm]{\Large
    Technical Report no. 2004-02
  }
  \raisebox{5mm}[0mm][0mm]{
    \textbf{
      \begin{tabular}{c}
Lock-Free and Practical Deques using Single-Word \\Compare-And-Swap
     \end{tabular}
    }
  }
}
\author{\raisebox{-15mm}[0mm][0mm]{\textbf{\Large Håkan Sundell}} \and \raisebox{-15mm}[0mm][0mm]{\textbf{\Large Philippas Tsigas}}}
\date{
  \vspace{\stretch{1}}
  \enlargethispage{1.1\baselineskip}
  \includegraphics{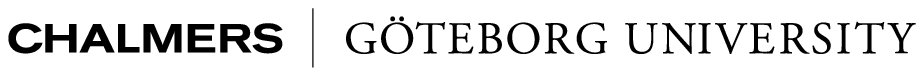} \\
  \vspace{5mm}
  \includegraphics{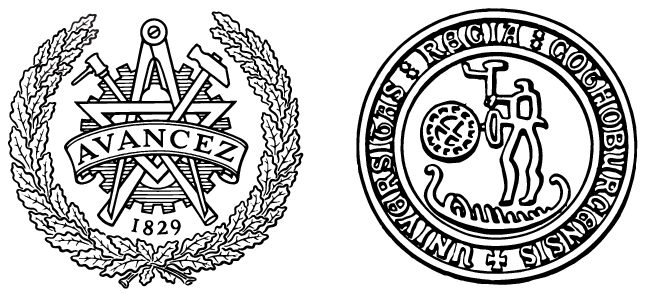} \\
  \vspace{12mm}
  Department of Computing Science \\
  Chalmers University of Technology
    and G\"{o}teborg University \\
  SE-412 96 G\"{o}teborg, Sweden \\
  \vspace{12mm}
  G\"{o}teborg, 2004
}    
\maketitle
\end{titlepage}

\newpage
\thispagestyle{empty}
\mbox{}
\vspace{\stretch{1}}

\noindent
{\large
  \begin{tabular}{l}
    \includegraphics{Logos/ChalmGUmarke.eps} \\[3ex]
    Technical Report in Computing Science at \\
    Chalmers University of Technology and G\"oteborg University
    \vspace{3ex} \\
    Technical Report no. 2004-02 \\
    ISSN: 1650-3023 
    \vspace{3ex} \\
    Department of Computing Science \\
    Chalmers University of Technology
      and G\"{o}teborg University \\
    SE-412 96 G\"{o}teborg, Sweden
    \vspace{3ex} \\
    G\"{o}teborg, Sweden, 2004
  \end{tabular}
}

\newpage

\twocolumn


\begin{abstract}
We present an efficient and practical lock-free implementation of a concurrent deque that is disjoint-parallel accessible and uses atomic primitives which are available in modern computer systems. Previously known lock-free algorithms of deques are either based on non-available atomic synchronization primitives, only implement a subset of the functionality, or are not designed for disjoint accesses. Our algorithm is based on a doubly linked list, and only requires single-word compare-and-swap atomic primitives, even for dynamic memory sizes. We have performed an empirical study using full implementations of the most efficient algorithms of lock-free deques known. For systems with low concurrency, the algorithm by Michael shows the best performance. However, as our algorithm is designed for disjoint accesses, it performs significantly better on systems with high concurrency and non-uniform memory architecture.
\end{abstract}

\section{Introduction}\label{sect.intro}

A deque (i.e. double-ended queue) is a fundamental data structure. For example, deques are often used for implementing the ready queue used for scheduling of tasks in operating systems. A deque supports four operations, the \textit{PushRight}, the \textit{PopRight}, the \textit{PushLeft}, and the \textit{PopLeft} operation. The abstract definition of a deque is a list of values, where the \textit{PushRight/PushLeft} operation adds a new value to the right/left edge of the list. The \textit{PopRight/PopLeft} operation correspondingly removes and returns the value on the right/left edge of the list.

To ensure consistency of a shared data object in a concurrent environment, the most common method is mutual exclusion, i.e. some form of locking. Mutual exclusion degrades the system's overall performance \cite{SilG94} as it causes blocking, i.e. other concurrent operations can not make any progress while the access to the shared resource is blocked by the lock. Mutual exclusion can also cause deadlocks, priority inversion and even starvation.

Researchers have addressed these problems by proposing non-blocking algorithms for shared data objects. Non-blocking methods do not involve mutual exclusion, and therefore do not suffer from the problems that blocking could generate. Lock-free implementations are non-blocking and guarantee that regardless of the contention caused by concurrent operations and the interleaving of their sub-operations, always at least one operation will progress. However, there is a risk for starvation as the progress of some operations could cause some other operations to never finish. Wait-free \cite{Her91} algorithms are lock-free and moreover they avoid starvation as well, as all operations are then guaranteed to finish in a limited number of their own steps. Recently, researchers also include obstruction-free \cite{HerLM03} implementations to be non-blocking, although this kind of implementation is weaker than lock-free and thus does not guarantee progress of any concurrent operation.

The implementation of a lock-based concurrent deque is a trivial task, and can preferably be constructed using either a doubly linked list or a cyclic array, protected with either a single lock or with multiple locks where each lock protects a part of the shared data structure. To the best of our knowledge, there exists no implementations of wait-free deques, but several lock-free implementations have been proposed. However, all previously lock-free deques lack in several important aspects, as they either only implement a subset of the operations that are normally associated with a deque and have concurrency restrictions\footnote{The algorithm by Arora et al does not support push operations on both ends, and does not allow concurrent invocations of the push operation and a pop operation on the opposite end.} like Arora et al \cite{AroBP98}, or are based on atomic hardware primitives like Double-Word Compare-And-Swap (CAS2)\footnote{A CAS2 operations can atomically read-and-possibly-update the contents of two non-adjacent memory words. This operation is also sometimes called DCAS in the literature.} which is not available in modern computer systems. Greenwald \cite{Gre99} presented a CAS2-based deque implementation, and there is also a publication series of a CAS2-based deque implementation \cite{ADFGMSS00},\cite{DFGMSS00} with the latest version by Martin et al \cite{MarMS02}. Independently of our work, Michael \cite{Mic03} has developed a deque implementation based on Compare-And-Swap (CAS)\footnote{The standard CAS operation can atomically read-and-possibly-update the contents of a single memory word}. However, it is not disjoint-parallel accessible as all operations have to synchronize, even though they operate on different ends of the deque. Secondly, in order to use dynamic maximum deque sizes it requires an extended CAS-operation that can atomically operate on two adjacent words, which is not available\footnote{It is available on the Intel IA-32, but not on the Sparc or MIPS microprocessor architectures.} on all modern platforms.

In this paper we present a lock-free algorithm of a concurrent deque that is disjoint-parallel accessible\footnote{There is a general and formal definition called disjoint-access-parallel by Israeli and Rappoport \cite{IsrR94}} (in the sense that operations on different ends of the deque do not necessarily interfere with each other) and implemented using common synchronization primitives that are available in modern systems. It can be extended to use dynamic maximum deque sizes (in the presence of a lock-free dynamic memory handler), still using normal CAS-operations. The algorithm is described in detail later in this paper, and the aspects concerning the underlying lock-free memory management are also presented. The precise semantics of the operations are defined and we give a proof that our implementation is lock-free and linearizable \cite{HerW90}. 

We have performed experiments that compare the performance of our algorithm with two of the most efficient algorithms of lock-free deques known; \cite{Mic03} and \cite{MarMS02}, the latter implemented using results from \cite{DetMMS01} and \cite{HarFP02}. Experiments were performed on three different multiprocessor system equipped with either 2,4 or 29 processors. All of the systems are using different operating systems. Our results show that the CAS-based algorithms outperforms the CAS2-based implementations\footnote{The CAS2 operation was implemented in software, using either mutual exclusion or the results from \cite{HarFP02}, which presented an software CAS\textit{n} (CAS for \textit{n} non-adjacent words) implementation.} for any number of threads, and for the system with full concurrency and non-uniform memory architecture our algorithm performs significantly better than the algorithm in \cite{Mic03}.

The rest of the paper is organized as follows. In Section \ref{sect.system} we describe the type of systems that our implementation is aimed for. The actual algorithm is described in Section \ref{sect.algorithm}. In Section \ref{sect.correctness} we define the precise semantics for the operations on our implementation, and show their correctness by proving the lock-free and linearizability properties. The experimental evaluation is presented in Section \ref{sect.experiments}. We conclude the paper with Section \ref{sect.conclusions}.

\begin{figure}
\begin{center}
\psfig{figure=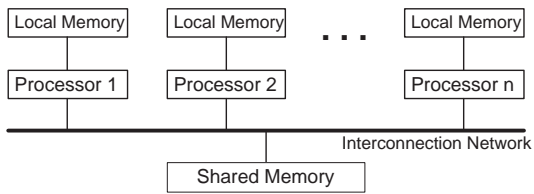, width=7cm, angle=0}
\end{center}
\caption{Shared Memory Multiprocessor System Structure}
\label{fig:structure}
\end{figure}


\section{System Description}\label{sect.system}

A typical abstraction of a shared memory multi-processor system configuration is depicted in Figure \ref{fig:structure}. Each node of the system contains a processor together with its local memory. All nodes are connected to the shared memory via an interconnection network. A set of co-operating tasks is running on the system performing their respective operations. Each task is sequentially executed on one of the processors, while each processor can serve (run) many tasks at a time. The co-operating tasks, possibly running on different processors, use shared data objects built in the shared memory to co-ordinate and communicate. Tasks synchronize their operations on the shared data objects through sub-operations on top of a cache-coherent shared memory. The shared memory may not though be uniformly accessible for all nodes in the system; processors can have different access times on different parts of the memory.


\section{Algorithm}\label{sect.algorithm}

The algorithm is based on a doubly-linked list data structure. To use the data structure as a deque, every node contains a value. The fields of each node item are described in Figure \ref{fig:algorithm1} as it is used in this implementation.

In order to make the deque construction concurrent and non-blocking, we are using three of the standard atomic synchronization primitives, Test-And-Set (TAS), Fetch-And-Add (FAA) and Compare-And-Swap (CAS). Figure \ref{fig:atomic_primitives} describes the specification of these primitives which are available in most modern platforms.

\begin{figure}
\begin{small}
\begin{tabbing}
\ \ \ \ \ \ \ \ \ \=
          \ \ \ \ \= 
               \ \ \ \ \= 
                    \ \ \ \ \=
                         \ \ \ \ \=
\\[-0.5cm]
\textbf{function} TAS(value:\textbf{pointer to word}):\textbf{boolean} \\
\> \textbf{atomic} \textbf{do} \\
\> \> \textbf{if} *value=0 \textbf{then} \\
\> \> \> *value:=1; \\
\> \> \> \textbf{return} \textbf{true}; \\
\> \> \textbf{else} \textbf{return} \textbf{false}; \\
\\
\textbf{procedure} FAA(address:\textbf{pointer to word}, number:\textbf{integer}) \\
\> \textbf{atomic} \textbf{do} \\
\> \> *address := *address + number; \\
\\
\textbf{function} CAS(address:\textbf{pointer to word}, oldvalue:\textbf{word},\\
~ newvalue:\textbf{word}):\textbf{boolean} \\
\> \textbf{atomic} \textbf{do} \\
\> \> \textbf{if} *address = oldvalue \textbf{then} \\
\> \> \> *address := newvalue; \\
\> \> \> \textbf{return} \textbf{true}; \\
\> \> \textbf{else} \textbf{return} \textbf{false};
\end{tabbing}
\end{small}
\caption{The Test-And-Set (TAS), Fetch-And-Add (FAA) and Compare-And-Swap (CAS) atomic primitives.}
\label{fig:atomic_primitives}
\end{figure}

\begin{figure}
\begin{center}
\psfig{figure=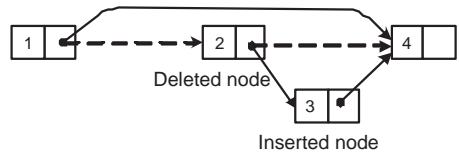, width=6cm, angle=0}
\end{center}
\caption{Concurrent insert and delete operation can delete both nodes.}
\label{fig:DeletedInsert}
\end{figure}

To insert or delete a node from the list we have to change the respective set of prev and next pointers. These have to be changed consistently, but not necessarily all at once. Our solution is to treat the doubly-linked list as being a singly-linked list with auxiliary information in the prev pointers, with the next pointers being updated before the prev pointers. Thus, the next pointers always form a consistent singly-linked list, but the prev pointers only give hints for where to find the previous node. This is possible because of the observation that a ``late'' non-updated prev pointer will always point to a node that is directly or some steps previous of the current node, and from that ``hint'' position it is always possible to traverse\footnote{As will be shown later, we have defined the deque data structure in a way that makes it possible to traverse even through deleted nodes, as long as they are referenced in some way.} through the next pointers to reach the directly previous node.

One problem, that is general for non-blocking implementations that are based on the singly-linked list structure, arises when inserting a new node into the list. Because of the linked-list structure one has to make sure that the previous node is not about to be deleted. If we are changing the next pointer of this previous node atomically with CAS, to point to the new node, and then immediately afterwards the previous node is deleted - then the new node will be deleted as well, as illustrated in Figure \ref{fig:DeletedInsert}. There are several solutions to this problem. One solution is to use the CAS2 operation as it can change two pointers atomically, but this operation is not available in any modern multiprocessor system. A second solution is to insert auxiliary nodes \cite{Val95} between every two normal nodes, and the latest method introduced by Harris \cite{Har01} is to use a deletion mark. This deletion mark is updated atomically together with the next pointer. Any concurrent insert operation will then be notified about the possibly set deletion mark, when its CAS operation will fail on updating the next pointer of the to-be-previous node. For our doubly-linked list we need to be informed also when inserting using the prev pointer. In order to be able to atomically update both the prev and the next pointer together with the deletion mark, all of these have to be put together in one memory word. For a 32-bit word this means a maximum of 32 768 (or 2 147 483 648 for a 64-bit word) possibly addressable nodes using the prev or next pointers. However, as will be shown later in Section \ref{sect.dynamicsize}, the algorithm can easily be extended to handle dynamic maximum sizes, thus making this limit obsolete.

\subsection{Memory Management}

As we are concurrently (with possible preemptions) traversing nodes that will be continuously allocated and reclaimed, we have to consider several aspects of memory management. No node should be reclaimed and then later re-allocated while some other process is (or will be) traversing that node. This can be solved for example by careful reference counting. We have selected the lock-free memory management scheme invented by Valois \cite{Val95} and corrected by Michael and Scott \cite{MicS95}, which makes use of the FAA and CAS atomic synchronization primitives. Using this scheme we can assure that a node can only be reclaimed when there is no prev or next pointer in the list that points to it. One problem with this scheme is that it can not handle cyclic garbage (i.e. 2 or more nodes that should be recycled but reference each other, and therefore each node keeps a positive reference count, although they are not referenced by the main structure). Our solution is to make sure to break potential cyclic references directly before a node is possibly recycled.

Another memory management issue is how to de-reference pointers safely. If we simply de-reference the pointer, it might be that the corresponding node has been reclaimed before we could access it. It can also be that the deletion mark that is connected to the prev or next pointer was set, thus marking that the node is deleted. The following functions are defined for safe handling of the memory management:

\

\begin{small}

\textbf{function} MALLOC\_NODE() :\textbf{pointer to} Node

\textbf{function} READ\_PREV(address:\textbf{pointer to} Link) :\textbf{pointer to} Node

\textbf{function} READ\_NEXT(address:\textbf{pointer to} Link) :\textbf{pointer to} Node

\textbf{function} READ\_PREV\_DEL(address:\textbf{pointer to} Link) :\textbf{pointer to} Node

\textbf{function} READ\_NEXT\_DEL(address:\textbf{pointer to} Link) :\textbf{pointer to} Node

\textbf{function} COPY\_NODE(node:\textbf{pointer to} Node) :\textbf{pointer to} Node

\textbf{procedure} RELEASE\_NODE(node:\textbf{pointer to} Node)

\end{small}
\

The function \textit{MALLOC\_NODE} allocates a new node from the memory pool of pre-allocated nodes. The function \textit{RELEASE\_NODE} decrements the reference counter on the corresponding given node. If the reference count reaches zero, the function then calls the \textit{ReleaseReferences} function that will recursively call \textit{RELEASE\_NODE} on the nodes that this node has owned pointers to, and then it reclaims the node. The function \textit{COPY\_NODE} increases the reference counter for the corresponding given node. \textit{READ\_PREV}, \textit{READ\_PREV\_DEL}, \textit{READ\_NEXT} and \textit{READ\_NEXT\_DEL} atomically de-references the given link in the corresponding direction and increases the reference counter for the corresponding node. In case the deletion mark of the link is set, the functions \textit{READ\_PREV} and \textit{READ\_NEXT} return NULL.

\begin{figure*}
\begin{minipage}[t]{5.0cm}
\begin{small}
\begin{tabbing}
\ \ \ \ \ \ \ \ \ \ \ \=
          \ \ \ \ \= 
               \ \ \ \ \= 
                    \ \ \ \ \=
                         \ \ \ \ \=\\[-0.5cm]
\textbf{union} Link\\
\> \_: \textbf{word}\\
\> $\langle prev,next,d \rangle$: $\langle$\textbf{pointer to} Node, \textbf{pointer to} Node, \textbf{boolean}$\rangle$\\
\\
\textbf{structure} Node\\
\> value: \textbf{pointer to word}\\
\> link: \textbf{union} Link\\
\\
// Global variables\\
head, tail: \textbf{pointer to} Node\\
// Local variables\\
node, prev, prev2, next, next2: \textbf{pointer to} Node\\
link1, link2, lastlink: \textbf{union} Link\\
\\
\textbf{function} CreateNode(value:\textbf{pointer to word}):\textbf{pointer to} Node\\
C1  \> node:=MALLOC\_NODE();\\
C2  \> node.value:=value;\\
C3  \> \textbf{return} node;\\
\\
\textbf{procedure} ReleaseReferences(node:\textbf{pointer to} Node)\\
RR1 \> \> RELEASE\_NODE(node.link.prev);\\
RR2 \> \> RELEASE\_NODE(node.link.next);\\
\\
\textbf{procedure} PushLeft(value: \textbf{pointer to word})\\
L1 \> node:=CreateNode(value);\\
L2 \> prev:=COPY\_NODE(head);\\
L3 \> next:=READ\_NEXT(\&prev.link);\\
L4 \> \textbf{while} \textbf{true} \textbf{do}\\
L5 \> \> link1:=prev.link;\\
L6 \> \> \textbf{if} link1.next $\neq$ next \textbf{then}\\
L7 \> \> \> RELEASE\_NODE(next);\\
L8 \> \> \> next:=READ\_NEXT(\&prev.link);\\
L9 \> \> \> \textbf{continue};\\
L10 \> \> node.link:=$\langle$prev,link1.next,\textbf{false}$\rangle$;\\
L11 \> \> link2:=$\langle$link1.prev,node,\textbf{false}$\rangle$;\\
L12 \> \> \textbf{if} CAS(\&prev.link,link1,link2) \textbf{then}\\
L13 \> \> \> COPY\_NODE(node);\\
L14 \> \> \> \textbf{break};\\
L15 \> \> \textit{Back-Off}\\
L16 \> PushCommon(node,next);\\
\\
\textbf{procedure} PushRight(value: \textbf{pointer to word})\\
R1 \> node:=CreateNode(value);\\
R2 \> next:=COPY\_NODE(tail);\\
R3 \> prev:=READ\_PREV(\&next.link);\\
R4 \> \textbf{while} \textbf{true} \textbf{do}\\
R5 \> \> link1:=prev.link;\\
R6 \> \> \textbf{if} link1.next $\neq$ next \textbf{or} prevlink.d $=$ \textbf{true} \textbf{then}\\
R7 \> \> \> prev:=HelpInsert(prev,next);\\
R8 \> \> \> \textbf{continue};\\
R9 \> \> node.link:=$\langle$prev,link1.next,\textbf{false}$\rangle$;\\
R10 \> \> link2:=$\langle$link1.prev,node,\textbf{false}$\rangle$;\\
R11 \> \> \textbf{if} CAS(\&prev.link,link1,link2) \textbf{then}\\
R12 \> \> \> COPY\_NODE(node);\\
R13 \> \> \> \textbf{break};\\
R14 \> \> \textit{Back-Off}\\
R15 \> PushCommon(node,next);
\end{tabbing}
\end{small}
\end{minipage}
\hfill
\begin{minipage}[t]{5.0cm}
\begin{small}
\begin{tabbing}
\ \ \ \ \ \ \ \ \ \ \ \=
          \ \ \ \ \= 
               \ \ \ \ \= 
                    \ \ \ \ \=
                         \ \ \ \ \=\\[-0.5cm]
\textbf{procedure} PushCommon(node: \textbf{pointer to} Node, next: \textbf{pointer to} Node)\\
P1 \> \textbf{while} \textbf{true} \textbf{do}\\
P2 \> \> link1:=next.link;\\
P3 \> \> link2:=$\langle$node,link1.next,\textbf{false}$\rangle$;\\
P4 \> \> \textbf{if} link1.d $=$ \textbf{true} \textbf{or} node.link.d $=$ \textbf{true}\\
P5 \> \> ~ \textbf{or} node.link.next $\neq$ next \textbf{then}\\
P6 \> \> \> \textbf{break};\\
P7 \> \> \textbf{if} CAS(\&next.link,link1,link2) \textbf{then}\\
P8 \> \> \> COPY\_NODE(node);\\
P9 \> \> \> RELEASE\_NODE(link1.prev);\\
P10 \> \> \> \textbf{if} node.link.d = \textbf{true} \textbf{then}\\
P11 \> \> \> \> prev2:=COPY\_NODE(node);\\
P12 \> \> \> \> prev2:=HelpInsert(prev2,next);\\
P13 \> \> \> \> RELEASE\_NODE(prev2);\\
P14 \> \> \> \textbf{break};\\
P15 \> \> \textit{Back-Off}\\
P16 \> RELEASE\_NODE(next);\\
P17 \> RELEASE\_NODE(node);\\
\\
\textbf{function} PopLeft(): \textbf{pointer to word}\\
PL1 \> prev:=COPY\_NODE(head);\\
PL2 \> \textbf{while} \textbf{true} \textbf{do}\\
PL3 \> \> node:=READ\_NEXT(\&prev.link);\\
PL4 \> \> \textbf{if} node $=$ tail \textbf{then} \\
PL5 \> \> \> RELEASE\_NODE(node);\\
PL6 \> \> \> RELEASE\_NODE(prev);\\
PL7 \> \> \> \textbf{return} $\bot$;\\
PL8 \> \> link1:=node.link;\\
PL9 \> \> \textbf{if} link1.d $=$ \textbf{true} \textbf{then}\\
PL10 \> \> \> DeleteNext(node);\\
PL11 \> \> \> RELEASE\_NODE(node);\\
PL12 \> \> \> \textbf{continue};\\
PL13 \> \> next:=COPY\_NODE(link1.next);\\
PL14 \> \> link2:=$\langle$link1.prev,link1.next,\textbf{true}$\rangle$\\
PL15 \> \> \textbf{if} CAS(\&node.link,link1,link2) \textbf{then}\\
PL16 \> \> \> DeleteNext(node);\\
PL17 \> \> \> prev:=HelpInsert(prev,next);\\
PL18 \> \> \> RELEASE\_NODE(prev);\\
PL19 \> \> \> RELEASE\_NODE(next);\\
PL20 \> \> \> value:=node.value;\\
PL21 \> \> \> \textbf{break};\\
PL22 \> \> RELEASE\_NODE(node);\\
PL23 \> \> RELEASE\_NODE(next);\\
PL24 \> \> \textit{Back-Off}\\
PL25 \> RemoveCrossReference(node);\\
PL26 \> RELEASE\_NODE(node);\\
PL27 \> \textbf{return} value;\\
\\
\textbf{function} PopRight(): \textbf{pointer to word}\\
PR1 \> next:=COPY\_NODE(tail);\\
PR2 \> \textbf{while} \textbf{true} \textbf{do}\\
PR3 \> \> node:=READ\_PREV(\&next.link);\\
PR4 \> \> link1:=node.link;\\
PR5 \> \> \textbf{if} link1.next $\neq$ next \textbf{or} link1.d = \textbf{true} \textbf{then}\\
PR6 \> \> \> node:=HelpInsert(node,next);\\
PR7 \> \> \> RELEASE\_NODE(node);\\
PR8 \> \> \> \textbf{continue};
\end{tabbing}
\end{small}
\end{minipage}
\caption{The algorithm, part 1(2).}
\label{fig:algorithm1}
\end{figure*}

\begin{figure*}
\begin{minipage}[t]{5.0cm}
\begin{small}
\begin{tabbing}
\ \ \ \ \ \ \ \ \ \ \ \=
          \ \ \ \ \= 
               \ \ \ \ \= 
                    \ \ \ \ \=
                         \ \ \ \ \=\\[-0.5cm]
PR9 \> \> \textbf{if} node $=$ head \textbf{then} \\
PR10 \> \> \> RELEASE\_NODE(next);\\
PR11 \> \> \> RELEASE\_NODE(node);\\
PR12 \> \> \> \textbf{return} $\bot$;\\
PR13 \> \> prev:=COPY\_NODE(link1.prev);\\
PR14 \> \> link2:=$\langle$link1.prev,link1.next,\textbf{true}$\rangle$\\
PR15 \> \> \textbf{if} CAS(\&node.link,link1,link2) \textbf{then}\\
PR16 \> \> \> DeleteNext(node);\\
PR17 \> \> \> prev:=HelpInsert(prev,next);\\
PR18 \> \> \> RELEASE\_NODE(prev);\\
PR19 \> \> \> RELEASE\_NODE(next);\\
PR20 \> \> \> value:=node.value;\\
PR21 \> \> \> \textbf{break};\\
PR22 \> \> RELEASE\_NODE(prev);\\
PR23 \> \> RELEASE\_NODE(node);\\
PR24 \> \> \textit{Back-Off}\\
PR25 \> RemoveCrossReference(node);\\
PR26 \> RELEASE\_NODE(node);\\
PR27 \> \textbf{return} value;\\
\\
\textbf{procedure} DeleteNext(node: \textbf{pointer to} Node)\\
DN1 \> lastlink.d:=\textbf{true};\\
DN2 \> prev:=READ\_PREV\_DEL(\&node.link);\\
DN3 \> next:=READ\_NEXT\_DEL(\&node.link);\\
DN4 \> \textbf{while} \textbf{true} \textbf{do}\\
DN5 \> \> \textbf{if} prev $=$ next \textbf{then} \textbf{break};\\
DN6 \> \> \textbf{if} next.link.d = \textbf{true} \textbf{then}\\

DN7 \> \> \> next2:=READ\_NEXT\_DEL(\&next.link);\\
DN8 \> \> \> RELEASE\_NODE(next);\\
DN9 \> \> \> next:=next2;\\
DN10 \> \> \> \textbf{continue};\\
DN11 \> \> prev2:=READ\_NEXT(\&prev.link);\\
DN12 \> \> \textbf{if} prev2 $=$ NULL \textbf{then}\\

DN13 \> \> \> \textbf{if} lastlink.d = \textbf{false} \textbf{then}\\
DN14 \> \> \> \> DeleteNext(prev);\\
DN15 \> \> \> \> lastlink.d:=\textbf{true};\\
DN16 \> \> \> prev2:=READ\_PREV\_DEL(\&prev.link);\\
DN17 \> \> \> RELEASE\_NODE(prev);\\
DN18 \> \> \> prev:=prev2;\\
DN19 \> \> \> \textbf{continue};\\
DN20 \> \> link1:=$\langle$prev.link.prev,prev2,\textbf{false}$\rangle$;\\
DN21 \> \> \textbf{if} prev2 $\neq$ node \textbf{then}\\
DN22 \> \> \> lastlink.d:=\textbf{false};\\
DN23 \> \> \> RELEASE\_NODE(prev);\\
DN24 \> \> \> prev:=prev2;\\
DN25 \> \> \> \textbf{continue};\\
DN26 \> \> RELEASE\_NODE(prev2);\\
DN27 \> \> link2:=$\langle$link1.prev,node.link.next,\textbf{false}$\rangle$;\\
DN28 \> \> \textbf{if} CAS(\&prev.link,link1,link2) \textbf{then}\\
DN29 \> \> \> COPY\_NODE(link2.next);\\
DN30 \> \> \> RELEASE\_NODE(node);\\
DN31 \> \> \> \textbf{break};\\
DN32 \> \> \textit{Back-Off}\\
\end{tabbing}
\end{small}
\end{minipage}
\hfill
\begin{minipage}[t]{5.0cm}
\begin{small}
\begin{tabbing}
\ \ \ \ \ \ \ \ \ \ \ \=
          \ \ \ \ \= 
               \ \ \ \ \= 
                    \ \ \ \ \=
                         \ \ \ \ \=\\[-0.5cm]
DN33 \> RELEASE\_NODE(prev);\\
DN34 \> RELEASE\_NODE(next);\\
\\
\textbf{function} HelpInsert(prev: \textbf{pointer to} Node, node: \textbf{pointer to} Node)\\
~ :\textbf{pointer to} Node\\
HI1\> lastlink.d:=\textbf{true};\\
HI2\> \textbf{while} \textbf{true} \textbf{do}\\
HI3\> \> prev2:=READ\_NEXT(\&prev.link);\\
HI4\> \> \textbf{if} prev2 $=$ NULL \textbf{then}\\
HI5\> \> \> \textbf{if} lastlink.d = \textbf{false} \textbf{then}\\
HI6\> \> \> \> DeleteNext(prev);\\
HI7\> \> \> \> lastlink.d:=\textbf{true};\\
HI8\> \> \> prev2:=READ\_PREV\_DEL(\&prev.link);\\
HI9\> \> \> RELEASE\_NODE(prev);\\
HI10\> \> \> prev:=prev2;\\
HI11\> \> \> \textbf{continue};\\
HI12\> \> link1:=node.link;\\
HI13\> \> \textbf{if} link1.d $=$ \textbf{true} \textbf{then}\\
HI14\> \> \> RELEASE\_NODE(prev2);\\
HI15\> \> \> \textbf{break};\\
HI16\> \> \textbf{if} prev2 $\neq$ node \textbf{then}\\
HI17\> \> \> lastlink.d:=\textbf{false};\\
HI18\> \> \> RELEASE\_NODE(prev);\\
HI19\> \> \> prev:=prev2;\\
HI20\> \> \> \textbf{continue};\\
HI21\> \> RELEASE\_NODE(prev2);\\
HI22\> \> link2:=$\langle$prev,link1.next,\textbf{false}$\rangle$;\\
HI23\> \> \textbf{if} CAS(\&node.link,link1,link2) \textbf{then}\\
HI24\> \> \> COPY\_NODE(prev);\\
HI25\> \> \> RELEASE\_NODE(link1.prev);\\
HI26\> \> \> \textbf{if} prev.link.d $=$ \textbf{true} \textbf{then} \textbf{continue};\\
HI27\> \> \> \textbf{break};\\
HI28\> \> \textit{Back-Off}\\
HI29\> \textbf{return} prev;\\
\\
\textbf{procedure} RemoveCrossReference(node: \textbf{pointer to} Node)\\
RC1\> \textbf{while} \textbf{true} \textbf{do}\\
RC2\> \> link1:=node.link;\\
RC3\> \> prev:=link1.prev;\\
RC4\> \> \textbf{if} prev.link.d = \textbf{true} \textbf{then}\\
RC5\> \> \> prev2:=READ\_PREV\_DEL(\&prev.link);\\
RC6\> \> \> node.link:=$\langle$prev2,link1.next,\textbf{true}$\rangle$;\\
RC7\> \> \> RELEASE\_NODE(prev);\\
RC8\> \> \> \textbf{continue};\\
RC9\> \> next:=link1.next;\\
RC10\> \> \textbf{if} next.link.d = \textbf{true} \textbf{then}\\
RC11\> \> \> next2:=READ\_NEXT\_DEL(\&next.link);\\
RC12\> \> \> node.link:=$\langle$link1.prev,next2,\textbf{true}$\rangle$;\\
RC13\> \> \> RELEASE\_NODE(next);\\
RC14\> \> \> \textbf{continue};\\
RC15\> \> \textbf{break};\\
\end{tabbing}
\end{small}
\end{minipage}
\caption{The algorithm, part 2(2).}
\label{fig:algorithm2}
\end{figure*}

\subsection{Pushing and Popping Nodes}

The \textit{PushLeft} operation, see Figure \ref{fig:algorithm1}, first repeatingly tries in lines L4-L15 to insert the new node (\textit{node}) between the head node (\textit{prev}) and the leftmost node (\textit{next}), by atomically changing the next pointer of the head node. Before trying to update the link field, it assures in line L6 that the \textit{next} node is still the very next node of head, otherwise \textit{next} is updated in L7-L8. After the new node has been successfully inserted, it tries in lines P2-P15 to update the prev pointer of the next node. It retries until either i) it succeeds with the update, ii) it detects that either the next or new node is deleted, or iii) the next node is no longer directly next of the new node. In any of the two latter, the changes are due to concurrent Pop or Push operations, and the responsibility to update the prev pointer is then left to those. If the update succeeds, there is though the possibility that the new node was deleted (and thus the prev pointer of the \textit{next} node was possibly already updated by the concurrent Pop operation) directly before the CAS in line P7, and then the prev pointer is updated by calling the \textit{HelpInsert} function in line P12.

The \textit{PushRight} operation, see Figure \ref{fig:algorithm1}, first repeatedly tries in lines R4-R14 to insert the new node (\textit{node}) between the rightmost node (\textit{prev} and the tail node (\textit{next}), by atomically changing the next pointer of the \textit{prev} node. Before trying to update the link field, it assures in line R6 that the \textit{next} node is still the very next node of \textit{prev}, otherwise \textit{prev} is updated by calling the \textit{HelpInsert} function in R7-R8, which updates the prev pointer of the \textit{next} node. After the new node has been successfully inserted, it tries in lines P2-P15 to update the prev pointer of the next node, following the same scheme as for the \textit{PushLeft} operation.

The \textit{PopLeft} operation, see Figure \ref{fig:algorithm1}, first repeatedly tries in lines PL2-PL24 to mark the leftmost node (\textit{node}) as deleted. Before trying to update the link field, it first assures in line PL4 that the deque is not empty, and secondly in line PL9 that the node is not already marked for deletion. If the deque was detected to be empty, the function returns. If \textit{node} was marked for deletion, it tries to update the next pointer of the \textit{prev} node by calling the \textit{DeleteNext} function, and then \textit{node} is updated to be the leftmost node. If the prev pointer of \textit{node} was incorrect, it tries to update it by calling the \textit{HelpInsert} function. After the node has been successfully marked by the successful CAS operation in line PL15, it tries in line PL16 to update the next pointer of the \textit{prev} node by calling the \textit{DeleteNext} function, and in line PL17 to update the prev pointer of the \textit{next} node by calling the \textit{HelpInsert} function. After this, it tries in line PL25 to break possible cyclic references that includes \textit{node} by calling the \textit{RemoveCrossReference} function.

The \textit{PopRight} operation, see Figure \ref{fig:algorithm1}, first repeatedly tries in lines PR2-PR24 to mark the rightmost node (\textit{node}) as deleted. Before trying to update the link field, it assures i) in line PR5 that the node is not already marked for deletion, ii) in the same line that the prev pointer of the tail (\textit{next}) node is correct, and iii) in line PR9 that the deque is not empty. If the deque was detected to be empty, the function returns. If \textit{node} was marked for deletion or the prev pointer of the \textit{next} node was incorrect, it tries to update the prev pointer of the \textit{next} node by calling the \textit{HelpInsert} function, and then \textit{node} is updated to be the rightmost node. After the node has been successfully marked it follows the same scheme as the \textit{PopLeft} operation.

\subsection{Helping and Back-Off}

The \textit{DeleteNext} procedure, see Figure \ref{fig:algorithm2}, repeatedly tries in lines DN4-DN32 to delete (in the sense of a chain of next pointers starting from the head node) the given marked node (\textit{node}) by changing the next pointer from the previous non-marked node. First, we can safely assume that the next pointer of the marked node is always referring to a node (\textit{next}) to the right and the prev pointer is always referring to a node (\textit{prev}) to the left (not necessarily the first). Before trying to update the link field with the CAS operation in line DN28, it assures in line DN5 that \textit{node} is not already deleted, in line DN6 that the \textit{next} node is not marked, in line DN12 that the \textit{prev} node is not marked, and in DN21 that \textit{prev} is the previous node of \textit{node}. If \textit{next} is marked, it is updated to be the next node. If \textit{prev} is marked we might need to delete it before we can update \textit{prev} to one of its previous nodes and proceed with the current deletion, but in order to avoid infinite recursion, \textit{DeleteNext} is only called if a next pointer from a non-marked node to \textit{prev} has been observed (i.e. \textit{lastlink.d} is false). Otherwise if \textit{prev} is not the previous node of \textit{node} it is updated to be the next node.

The \textit{HelpInsert} procedure, see Figure \ref{fig:algorithm2}, repeatedly tries in lines HI2-HI28 to correct the prev pointer of the given node (\textit{node}), given a suggestion of a previous (not necessarily the first) node (\textit{prev}). Before trying to update the link field with the CAS operation in line HI23, it assures in line HI4 that the \textit{prev} node is not marked, in line HI13 that \textit{node} is not marked, and in line HI16 that \textit{prev} is the previous node of \textit{node}. If \textit{prev} is marked we might need to delete it before we can update \textit{prev} to one of its previous nodes and proceed with the current insertion, but in order to avoid unnecessary recursion, \textit{DeleteNext} is only called if a next pointer from a non-marked node to \textit{prev} has been observed (i.e. \textit{lastlink.d} is false). If \textit{node} is marked, the procedure is aborted. Otherwise if \textit{prev} is not the previous node of \textit{node} it is updated to be the next node. If the update in line HI23 succeeds, there is though the possibility that the \textit{prev} node was deleted (and thus the prev pointer of \textit{node} was possibly already updated by the concurrent Pop operation) directly before the CAS operation. This is detected in line HI26 and then the update is possibly retried with a new \textit{prev} node.

The \textit{RemoveCrossReference} procedure, see Figure \ref{fig:algorithm2}, tries to break cross-references between the given node (\textit{node}) and any of the nodes that it references, by repeatedly updating the prev or next pointer as long as it references a marked node. First, we can safely assume that the link field of \textit{node} is not concurrently updated by any other operation. Before the procedure is finished, it assures in line RC4 that the previous node (\textit{prev}) is not marked, and in line RC10 that the next node (\textit{next}) is not marked. As long as \textit{prev} is marked it is traversed to the left, and as long as \textit{next} is marked it is traversed to the right, while continuously updating the link field of \textit{node} in lines RC6 or RC12. 

Because the \textit{DeleteNext} and \textit{HelpInsert} are often used in the algorithm for ``helping'' late operations that might otherwise stop progress of other concurrent operations, the algorithm is suitable for pre-emptive as well as fully concurrent systems. In fully concurrent systems though, the helping strategy as well as heavy contention on atomic primitives, can downgrade the performance significantly. Therefore the algorithm, after a number of consecutive failed CAS operations (i.e. failed attempts to help concurrent operations) puts the current operation into back-off mode. When in back-off mode, the thread does nothing for a while, and in this way avoids disturbing the concurrent operations that might otherwise progress slower. The duration of the back-off is proportional to the number of threads, and for each consecutive entering of the back-off mode during one operation invocation, the duration of the back-off is increased exponentially.

\subsection{Extending to dynamic maximum sizes}\label{sect.dynamicsize}

In order to allow usage of a system-wide dynamic memory handler (which should be lock-free and have garbage collection capabilities), all significant bits of an arbitrary pointer value must be possible to be represented in both the next and prev pointers. In order to atomically update both the next and prev pointer together with the deletion mark, the CAS-operation would need the capability of atomically updating at least $30+30+1=61$ bits on a 32-bit system (and $62+62+1=125$ bits on a 64-bit system as the pointers are then 64 bit). However, most current 32 and 64-bit systems only support CAS-operations of single word-size.

An interesting observation of the current algorithm is that it never changes both the prev and next pointer in the atomic updates, and the pre-condition associated with the atomic CAS-update only involves the pointer that is changed. 

Therefore it is possible to keep the prev and next pointers in separate words, duplicating the deletion mark in each of the words. Thus, full pointer values can be used, still by only using standard CAS-operations. In order to preserve the correctness of the algorithm, the deletion mark of the next pointer should always be set first, in the \textit{PopLeft/PopRight} functions, and the deletion mark of the prev pointer should be possibly set in the very beginning of the \textit{DeleteNext} procedure. The remaining changes are trivial and the full extended algorithm is presented in Appendix \ref{sect.dynamic_algorithm}.


\section{Correctness}\label{sect.correctness}

In this section we present the proof of our algorithm. We first prove that our algorithm is a linearizable one \cite{HerW90} and then we prove that it is lock-free. A set of definitions that will help us to structure and shorten the proof is first explained in this section. We start by defining the sequential semantics of our operations and then introduce two definitions concerning concurrency aspects in general.

\begin{definition}
We denote with $Q_t$ the abstract internal state of a deque at the time $t$. $Q_t=[v_1,\dots,v_n]$ is viewed as an list of values $v$, where $|Q_t| \geq 0$. The operations that can be performed on the deque are \textit{PushLeft}($L$), \textit{PushRight}($R$), \textit{PopLeft}($PL$) and \textit{PopRight}($PR$). The time $t_1$ is defined as the time just before the atomic execution of the operation that we are looking at, and the time $t_2$ is defined as the time just after the atomic execution of the same operation. In the following expressions that define the sequential semantics of our operations, the syntax is $S_1:O_1,S_2$, where $S_1$ is the conditional state before the operation $O_1$, and $S_2$ is the resulting state after performing the corresponding operation:
\end{definition}

\begin{equation}
Q_{t_1}: {\bf L(v_1)} , Q_{t_2}=[v_1]+Q_{t_1}
\label{e:pushleft_seq_1}
\end{equation}

\begin{equation}
Q_{t_1}: {\bf R(v_1)} , Q_{t_2}=Q_{t_1}+[v_1]
\label{e:pushright_seq_1}
\end{equation}

\begin{equation}
Q_{t_1}=\emptyset: {\bf PL()=\bot} , Q_{t_2}=\emptyset
\label{e:popleft_seq_1}
\end{equation}

\begin{equation}
Q_{t_1}=[v_1]+Q_1: {\bf PL()=v_1} , Q_{t_2}=Q_1
\label{e:popleft_seq_2}
\end{equation}

\begin{equation}
Q_{t_1}=\emptyset: {\bf PR()=\bot} , Q_{t_2}=\emptyset
\label{e:popright_seq_1}
\end{equation}

\begin{equation}
Q_{t_1}=Q_1+[v_1]: {\bf PR()=v_1} , Q_{t_2}=Q_1
\label{e:popright_seq_2}
\end{equation}

\begin{definition}
In a global time model each concurrent operation $Op$ ``occupies" a time interval $[b_{Op}, f_{Op}]$ on the linear time axis $(b_{Op} < f_{Op})$. The precedence relation (denoted by `$\rightarrow$') is a relation that relates operations of a possible execution, $Op_1 \rightarrow Op_2$ means that $Op_1$ ends before $Op_2$ starts.
The precedence relation is a strict partial order. Operations incomparable under $\rightarrow$ are called {\em overlapping}. The overlapping relation is denoted by $\parallel$ and is commutative, i.e. $Op_1 \parallel Op_2$ and $Op_2 \parallel Op_1$. The precedence relation is extended to relate sub-operations of operations. Consequently, if $Op_1 \rightarrow Op_2$, then for any sub-operations $op_1$ and $op_2$ of $Op_1$ and $Op_2$, respectively, it holds that $op_1 \rightarrow op_2$. We also define the direct precedence relation ${\rightarrow}_d$, such that if $Op_1 {\rightarrow}_d Op_2$, then $Op_1 \rightarrow Op_2$ and moreover there exists no operation $Op_3$ such that $Op_1 \rightarrow Op_3 \rightarrow Op_2$.
\end{definition}

\begin{definition}
In order for an implementation of a shared concurrent data object to be linearizable \cite{HerW90}, for every concurrent execution there should exist an equal (in the sense of the effect) and valid (i.e. it should respect the semantics of the shared data object) sequential execution that respects the partial order of the operations in the concurrent execution.
\label{d:linearizable}
\end{definition}

Next we are going to study the possible concurrent executions of our implementation. First we need to define the interpretation of the abstract internal state of our implementation.

\begin{definition}
The value $v$ is \textit{present} ($\exists i . Q[i]=v$) in the abstract internal state $Q$ of our implementation, when there is a connected chain of next pointers (i.e. prev.link.next) from a \textit{present} node (or the head node) in the doubly linked list that connects to a node that contains the value $v$, and this node is not marked as deleted (i.e. node.link.d=false).
\label{d:abstract_state}
\end{definition}

\begin{definition}
The decision point of an operation is defined as the atomic statement where the result of the operation is finitely decided, i.e. independent of the result of any sub-operations after the decision point, the operation will have the same result. We define the state-read point of an operation to be the atomic statement where a sub-state of the priority queue is read, and this sub-state is the state on which the decision point depends. We also define the state-change point as the atomic statement where the operation changes the abstract internal state of the priority queue after it has passed the corresponding decision point.
\end{definition}

We will now use these points in order to show the existence and location in execution history of a point where the concurrent operation can be viewed as it occurred atomically, i.e. the \textit{linearizability point}.

\begin{lemma}
A \textit{PushRight} operation ($R(v)$), takes effect atomically at one statement.
\label{l:atomic_pushright1}
\end{lemma}
\begin{proof}
The decision, state-read and state-change point for a \textit{PushRight} operation which succeeds ($R(v)$), is when the CAS sub-operation in line R11 (see Figure \ref{fig:algorithm1}) succeeds. The state of the deque was ($Q_{t_1}=Q_1$) directly before the passing of the decision point. The prev node was the very last present node as it pointed (verified by R6 and the CAS in R11) to the tail node directly before the passing of the decision point. The state of the deque directly after passing the decision point will be $Q_{t_2}=Q_1+[v]$ as the next pointer of the prev node was changed to point to the new node which contains the value $v$. Consequently, the linearizability point will be the CAS sub-operation in line R11.
\end{proof}

\begin{lemma}
A \textit{PushLeft} operation ($L(v)$), takes effect atomically at one statement.
\label{l:atomic_pushleft1}
\end{lemma}
\begin{proof}
The decision, state-read and state-change point for a \textit{PushLeft} operation which succeeds ($L(v)$), is when the CAS sub-operation in line L12 (see Figure \ref{fig:algorithm1}) succeeds. The state of the deque was ($Q_{t_1}=Q_1$) directly before the passing of the decision point. The state of the deque directly after passing the decision point will be $Q_{t_2}=[v]+Q_1$ as the next pointer of the head node was changed to point to the new node which contains the value $v$. Consequently, the linearizability point will be the CAS sub-operation in line L12.
\end{proof}

\begin{lemma}
A \textit{PopRight} operation which fails ($PR()=\bot$), takes effect atomically at one statement.
\label{l:atomic_popright1}
\end{lemma}
\begin{proof}
The decision point for a \textit{PopRight} operation which fails ($PR()=\bot$) is the check in line PR9. Passing of the decision point together with the verification in line PR5 gives that the next pointer of the head node must have been pointing to the tail node ($Q_{t_1}=\emptyset$) directly before the read sub-operation of the link field in line PR4, i.e. the state-read point. Consequently, the linearizability point will be the read sub-operation in line PR4.
\end{proof}

\begin{lemma}
A \textit{PopRight} operation which succeeds ($PR()=v$), takes effect atomically at one statement.
\label{l:atomic_popright2}
\end{lemma}
\begin{proof}
The decision point for a \textit{PopRight} operation which succeeds ($PR()=v$) is when the CAS sub-operation in line PR15 succeeds. Passing of the decision point together with the verification in line PR5 gives that the next pointer of the to-be-deleted node must have been pointing to the tail node ($Q_{t_1}=Q_1+[v]$) directly before the CAS sub-operation in line PR15, i.e. the state-read point. Directly after passing the CAS sub-operation (i.e. the state-change point) the to-be-deleted node will be marked as deleted and therefore not present in the deque ($Q_{t_2}=Q_1$). Consequently, the linearizability point will be the CAS sub-operation in line PR15.
\end{proof}

\begin{lemma}
A \textit{PopLeft} operation which fails ($PL()=\bot$), takes effect atomically at one statement.
\label{l:atomic_popleft1}
\end{lemma}
\begin{proof}
The decision point for a \textit{PopLeft} operation which fails ($PL()=\bot$) is the check in line PL4. Passing of the decision point gives that the next pointer of the head node must have been pointing to the tail node ($Q_{t_1}=\emptyset$) directly before the read sub-operation of the link field in line PL3, i.e. the state-read point. Consequently, the linearizability point will be the read sub-operation in line PL3.
\end{proof}

\begin{lemma}
A \textit{PopLeft} operation which succeeds ($PL()=v$), takes effect atomically at one statement.
\label{l:atomic_popleft2}
\end{lemma}
\begin{proof}
The decision point for a \textit{PopLeft} operation which succeeds ($PL()=v$) is when the CAS sub-operation in line PL15 succeeds. Passing of the decision point together with the verification in line PL9 gives that the next pointer of the head node must have been pointing to the present to-be-deleted node ($Q_{t_1}=[v]+Q_1$) directly before the read sub-operation in line PL3, i.e. the state-read point. Directly after passing the CAS sub-operation in line PL15 (i.e. the state-change point) the to-be-deleted node will be marked as deleted and therefore not present in the deque ($\neg\exists i . Q_{t_2}[i]=v$). Unfortunately this does not match the semantic definition of the operation. 

However, none of the other concurrent operations linearizability points is dependent on the to-be-deleted node's state as marked or not marked during the time interval from the state-read to the state-change point. Clearly, the linearizability points of Lemmas \ref{l:atomic_pushright1} and \ref{l:atomic_pushleft1} are independent as the to-be-deleted node would be part (or not part if not present) of the corresponding $Q_1$ terms. The linearizability points of Lemmas \ref{l:atomic_popright1} and \ref{l:atomic_popleft1} are independent, as those linearizability points depend on the head node's next pointer pointing to the tail node or not. Finally, the linearizability points of Lemma \ref{l:atomic_popright2} as well as this lemma are independent, as the to-be-deleted node would be part (or not part if not present) of the corresponding $Q_1$ terms, otherwise the CAS sub-operation in line PL15 of this operation would have failed. 

Therefore all together, we could safely interpret the to-be-deleted node to be not present already directly after passing the state-read point (($Q_{t_2}=Q_1$). Consequently, the linearizability point will be the read sub-operation in line PL3.
\end{proof}

\begin{lemma}
When the deque is idle (i.e. no operations are being performed), all next pointers of present nodes are matched with a correct prev pointer from the corresponding present node (i.e. all linked nodes from the head or tail node are present in the deque).
\label{l:idle_allpresent}
\end{lemma}
\begin{proof}
We have to show that each operation takes responsibility for that the affected prev pointer will finally be correct after changing the corresponding next pointer. After successfully changing the next pointer in the \textit{PushLeft} (\textit{PushRight}) in line L12 (R11) operation, the corresponding prev pointer is tried to be changed in line P7 repeatedly until i) it either succeeds, ii) either the next or this node is deleted as detected in line P4, iii) or a new node is inserted as detected in line P5. If a new node is inserted the corresponding \textit{PushLeft} (\textit{PushRight}) operation will make sure that the prev pointer is corrected. If either the next or this node is deleted, the corresponding \textit{PopLeft} (\textit{PopRight}) operation will make sure that the prev pointer is corrected. If the prev pointer was successfully changed it is possible that this node was deleted before we changed the prev pointer of the next node. If this is detected in line P10, then the prev pointer of the next node is corrected by the \textit{HelpInsert} function.

After successfully marking the to-be-deleted nodes in line PL15 (PR15), the \textit{PopLeft} (\textit{PopRight}) functions will make sure that the connecting next pointer of the prev node will be changed to point to the closest present node to the right, by calling the \textit{DeleteNext} procedure in line PL16 (PR16). It will also make sure that the corresponding prev pointer of the next code will be corrected by calling the \textit{HelpInsert} function in line PL17 (PR17). 

The \textit{DeleteNext} procedure will repeatedly try to change the next pointer of the prev node that points to the deleted node, until it either succeeds changing the next pointer in line DN28 or some concurrent \textit{DeleteNext} already succeeded as detected in line DN5.

The \textit{HelpInsert} procedure will repeatedly try to change the prev pointer of the node to match with the next pointer of the prev node, until it either succeeds changing the prev pointer in line HI23 or the node is deleted as detected in line HI13. If it succeeded with changing the prev pointer, the prev node might have been deleted directly before changing the prev pointer, and therefore it is detected if the prev node is marked in line HI26 and then the procedure will continue trying to correctly change the prev pointer.
\end{proof}

\begin{lemma}
When the deque is idle, all previously deleted nodes are garbage collected.
\label{l:deleted_garbagecollect}
\end{lemma}
\begin{proof}
We have to show that each \textit{PopRight} or \textit{PopLeft} operation takes responsibility for that the deleted node will finally have no references to it. The possible references are caused by other nodes pointing to it. Following Lemma \ref{l:idle_allpresent} we know that no present nodes will reference the deleted node. It remains to show that all paths of references from a deleted node will finally reference a present node, i.e. there are no cyclic referencing. After the node is deleted in lines PL16 and PL17 (PR16 and PR17), it is assured by the \textit{PopLeft} (\textit{PopRight}) operation by calling the \textit{RemoveCrossReference} procedure in line PL25 (PR25) that both the next and prev pointers are pointing to a present node. If any of those present nodes are deleted before the referencing deleted node is garbage collected in line , the \textit{RemoveCrossReference} procedures called by the corresponding \textit{PopLeft} or \textit{PopRight} operation will assure that the next and prev pointers of the previously present node will point to present nodes, and so on recursively. The \textit{RemoveCrossReference} procedure repeatedly tries to change prev pointers to point to the previous node of the referenced node until the referenced node is present, detected in line RC4 and possibly changed in line RC6. The next pointer is correspondingly detected in line RC10 and possibly changed in line RC12.
\end{proof}

\begin{lemma}
The path of prev pointers from a node is always pointing a present node that is left of the current node.
\label{l:always_prev_left}
\end{lemma}
\begin{proof}
We will look at all possibilities where the prev pointer is set or changed. The setting in line L10 (R9) is clearly to the left as it is verified by L6 and L12 (R5 and R11). The change of the prev pointer in line P7 is to the left as verified by P5 and that nodes are never moved relatively to each other. The change of the prev pointer in line HI23 is to the left as verified by line HI3 and HI16. Finally, the change of the prev pointer in line RC6 is to the left as it is changed to the prev pointer of the previous node.
\end{proof}

\begin{lemma}
All operations will terminate if exposed to a limited number of concurrent changes to the deque.
\label{l:operation_terminate}
\end{lemma}
\begin{proof}
The amount of changes an operation could experience is limited. Because of the reference counting, none of the nodes which is referenced to by local variables can be garbage collected. When traversing through prev or next pointers, the memory management guarantees atomicity of the operations, thus no newly inserted or deleted nodes will be missed. We also know that the relative positions of nodes that are referenced to by local variables will not change as nodes are never moved in the deque. Most loops in the operations retry because a change in the state of some node(s) was detected in the ending CAS sub-operation, and then retry by re-reading the local variables (and possibly correcting the state of the nodes) until no concurrent changes was detected by the CAS sub-operation and therefore the CAS succeeded and the loop terminated. Those loops will clearly terminate after a limited number of concurrent changes. Included in that type of loops are L4-L15, R4-R14, P1-P15, PL2-PL24 and PR2-PR24.

The loop DN4-DN32 will terminate if either the prev node is equal to the next node in line DN5 or the CAS sub-operation in line DN28 succeeds. We know from the start of the execution of the loop, that the prev node is left of the to-be-deleted node which in turn is left of the next node. Following from Lemma \ref{l:always_prev_left} this order will hold by traversing the prev node through its prev pointer and traversing the next node through its next pointer. Consequently, traversing the prev node through the next pointer will finally cause the prev node to be directly left of the to-be-deleted node if this is not already deleted (and the CAS sub-operation in line DN28 will finally succeed), otherwise the prev node will finally be directly left of the next node (and in the next step the equality in line DN5 will hold). As long as the prev node is marked it will be traversed to the left in line DN16, and if it is the left-most marked node the prev node will be deleted by recursively calling \textit{DeleteNext} in line DN14. If the prev node is not marked it will be traversed to the right. As there is a limited number of changes and thus a limited number of marked nodes left of the to-be-deleted node, the prev node will finally traverse to the right and either of the termination criteria will be fulfilled.

The loop HI2-HI28 will terminate if either the to-be-corrected node is marked in line HI13 or if the CAS sub-operation in line HI23 succeeds and prev node is not marked. We know that from the start of the execution of the loop, that the prev node is left of the to-be-corrected node. Following from Lemma \ref{l:always_prev_left} this order will hold by traversing the prev node through its prev pointer. Consequently, traversing the prev node through the next pointer will finally cause the prev node to be directly left of the to-be-corrected node if this is not deleted (and the CAS sub-operation in line HI23 will finally succeed), otherwise line HI13 will succeed. As long as the prev node is marked it will be traversed to the left in line HI8, and if it is the left-most marked node the prev node will be deleted by calling \textit{DeleteNext} in line HI6. If the prev node is not marked it will be traversed to the right. As there is a limited number of changes and thus a limited number of marked nodes left of the to-be-corrected node, the prev node will finally traverse to the right and either of the termination criteria will be fulfilled.

The loop RC1-RC15 will terminate if both the prev node and the next node of the to-be-deleted node is not marked in line RC4 respectively line RC10. We know that from the start of the execution of the loop, the prev node is left of the to-be-deleted node and the next node is right of the to-be-deleted node. Following from Lemma \ref{l:always_prev_left}, traversing the prev node through the next pointer will finally reach a not marked node or the head node (which is not marked), and traversing the next node through the next pointer will finally reach a not marked node or the tail node (which is not marked), and both of the termination criteria will be fulfilled.
\end{proof}

\begin{lemma}
With respect to the retries caused by synchronization, one operation will always do progress regardless of the actions by the other concurrent operations.
\label{l:one_no_retry}
\end{lemma}
\begin{proof}
We now examine the possible execution paths of our implementation. There are several potentially unbounded loops that can delay the termination of the operations. We call these loops retry-loops. If we omit the conditions that are because of the operations semantics (i.e. searching for the correct criteria etc.), the loop retries when sub-operations detect that a shared variable has changed value. This is detected either by a subsequent read sub-operation or a failed CAS. These shared variables are only changed concurrently by other CAS sub-operations. According to the definition of CAS, for any number of concurrent CAS sub-operations, exactly one will succeed. This means that for any subsequent retry, there must be one CAS that succeeded. As this succeeding CAS will cause its retry loop to exit, and our implementation does not contain any cyclic dependencies between retry-loops that exit with CAS, this means that the corresponding \textit{PushRight}, \textit{PushLeft}, \textit{PopRight} or \textit{PopLeft} operation will progress. Consequently, independent of any number of concurrent operations, one operation will always progress.
\end{proof}

\begin{theorem}
The algorithm implements a correct, memory stable, lock-free and linearizable deque.
\end{theorem}
\begin{proof}
Following from Lemmas \ref{l:atomic_pushright1}, \ref{l:atomic_pushleft1}, \ref{l:atomic_popright1}, \ref{l:atomic_popright2}, \ref{l:atomic_popleft1} and \ref{l:atomic_popleft2} and by using the respective linearizability points, we can create an identical (with the same semantics) sequential execution that preserves the partial order of the operations in a concurrent execution. Following from Definition \ref{d:linearizable}, the implementation is therefore linearizable. 

Lemmas \ref{l:operation_terminate} and \ref{l:one_no_retry} give that our implementation is lock-free.

Following from Lemmas \ref{l:operation_terminate}, \ref{l:atomic_pushright1}, \ref{l:atomic_pushleft1}, \ref{l:atomic_popright1}, \ref{l:atomic_popright2}, \ref{l:atomic_popleft1} and \ref{l:atomic_popleft2} we can conclude that all operations will terminate with the correct result.

Following from Lemma \ref{l:deleted_garbagecollect} we know that the maximum memory usage will be proportional to the number of present values in the deque.

\end{proof}


\section{Experimental Evaluation}\label{sect.experiments}

\begin{figure*}
\begin{center}
\begin{tabular}{ll}
\psfig{figure=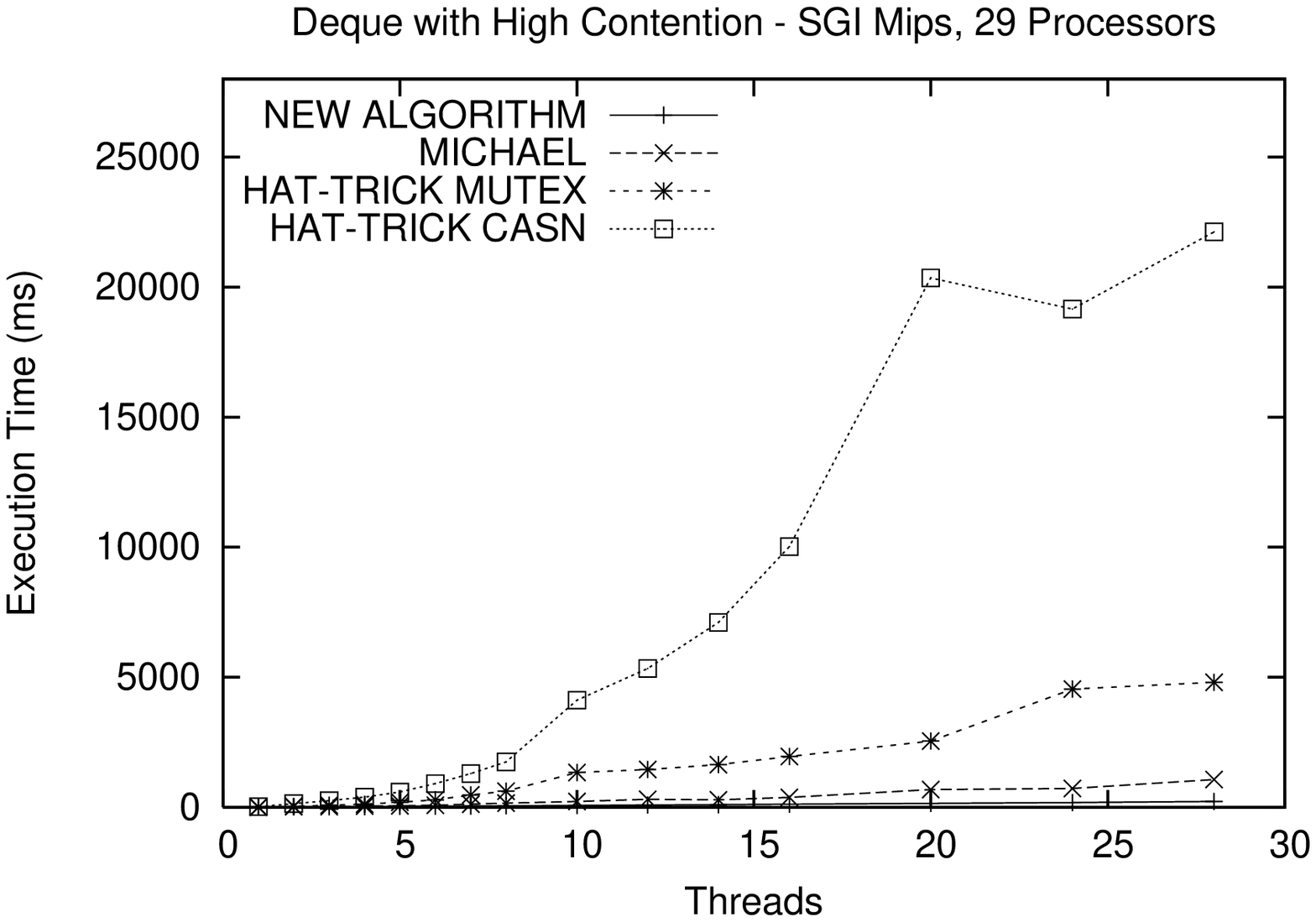, width=6.0cm, angle=270} &
\psfig{figure=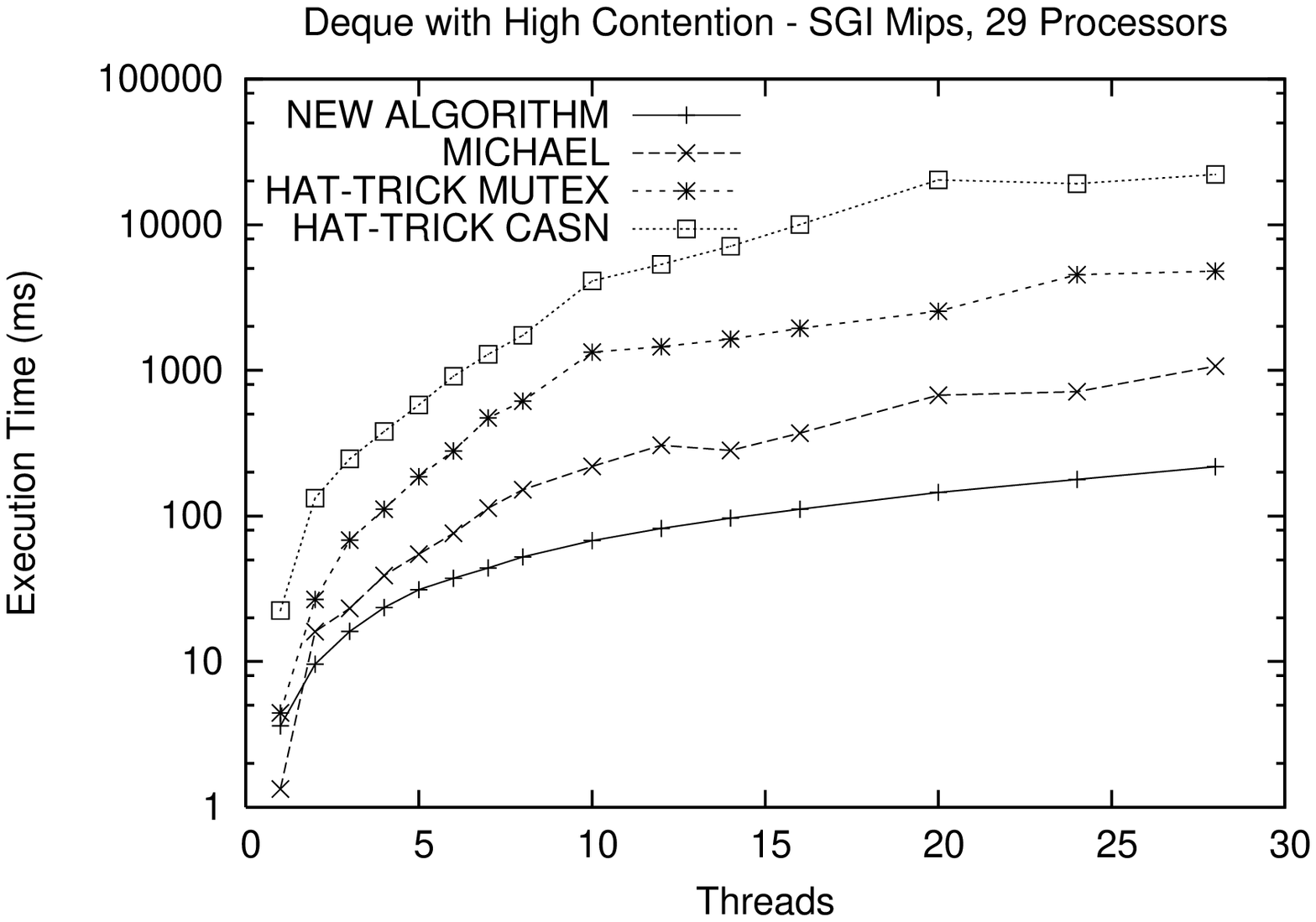, width=6.0cm, angle=270}\\
\psfig{figure=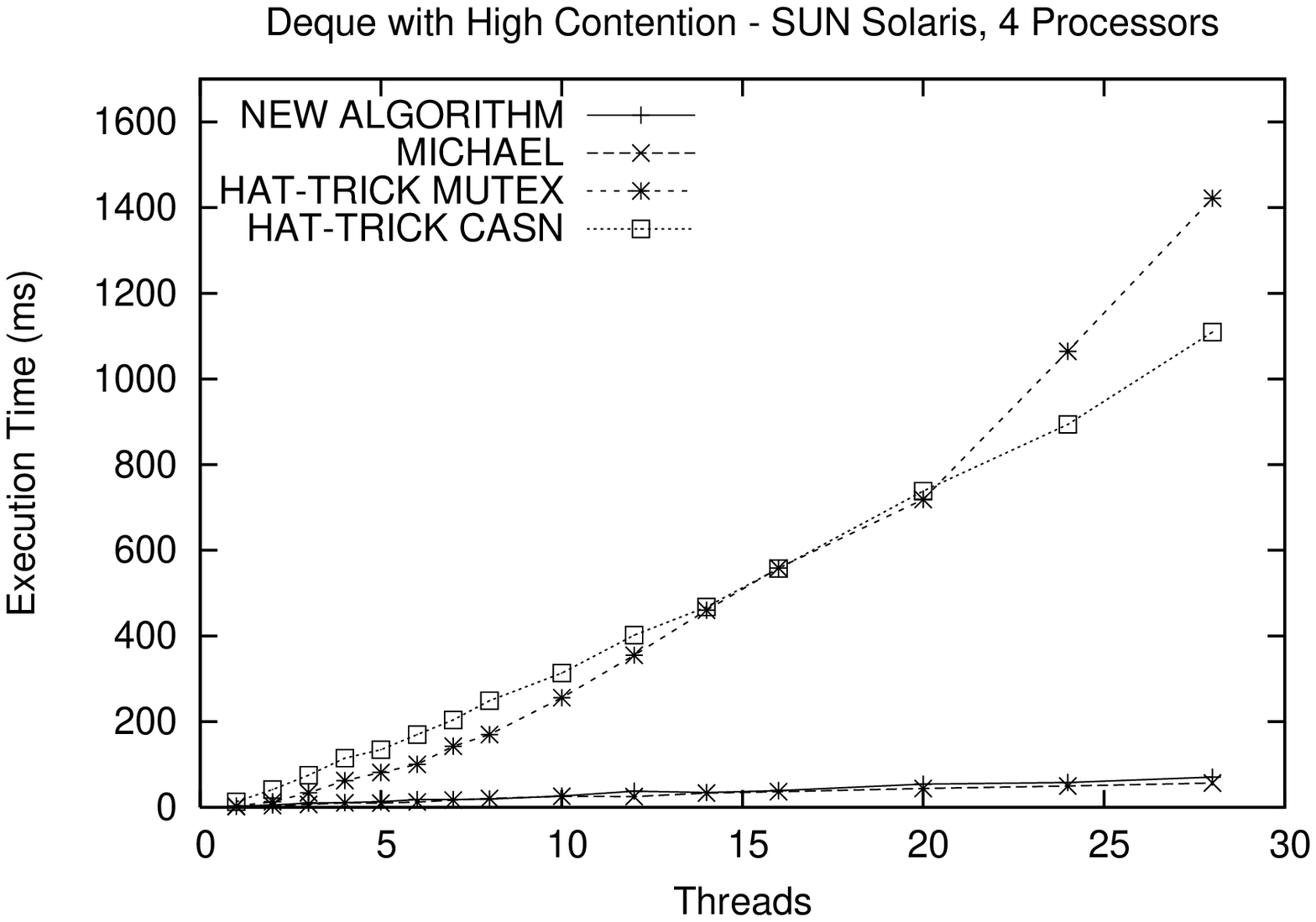, width=6.0cm, angle=270} &
\psfig{figure=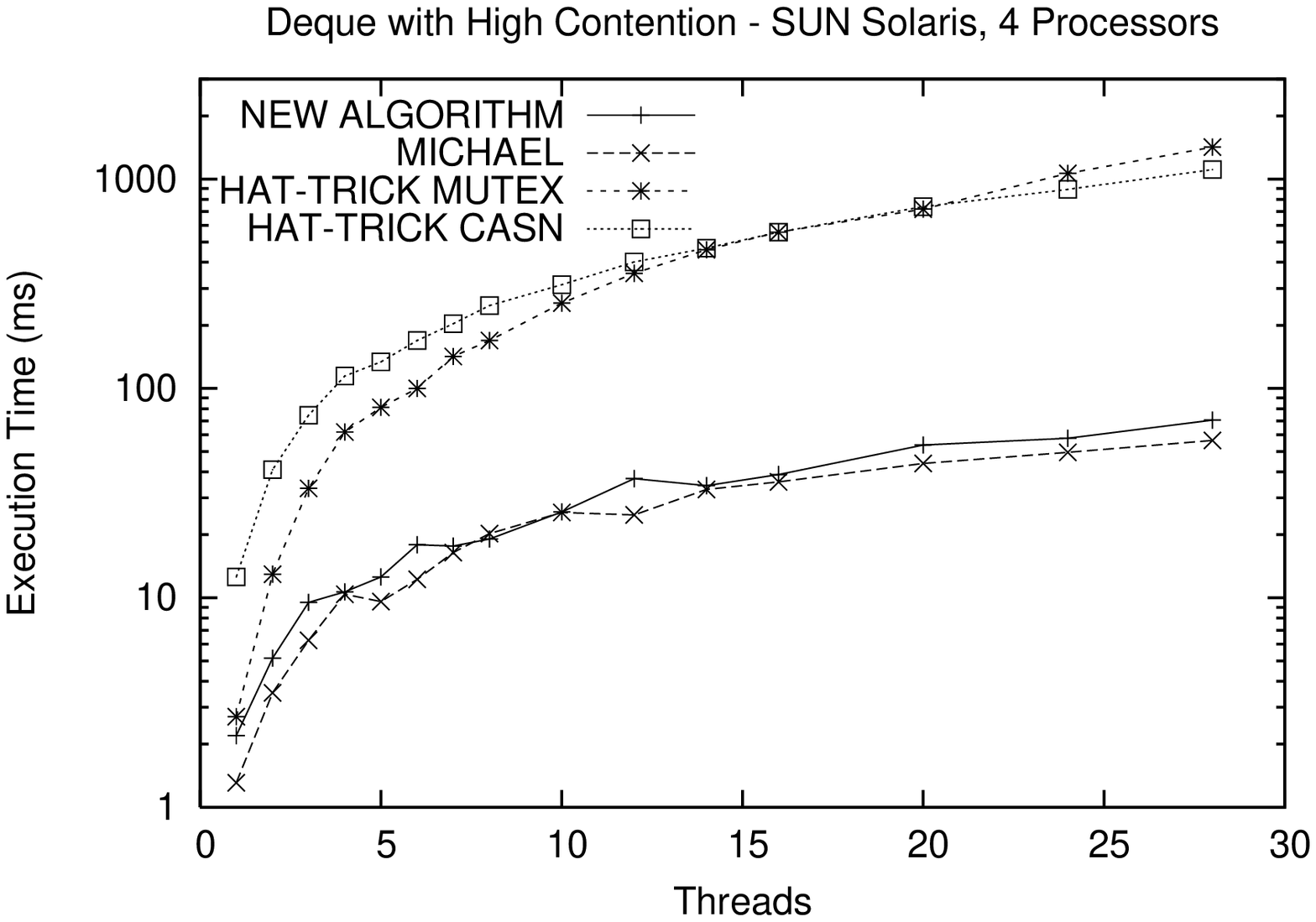, width=6.0cm, angle=270}\\
\psfig{figure=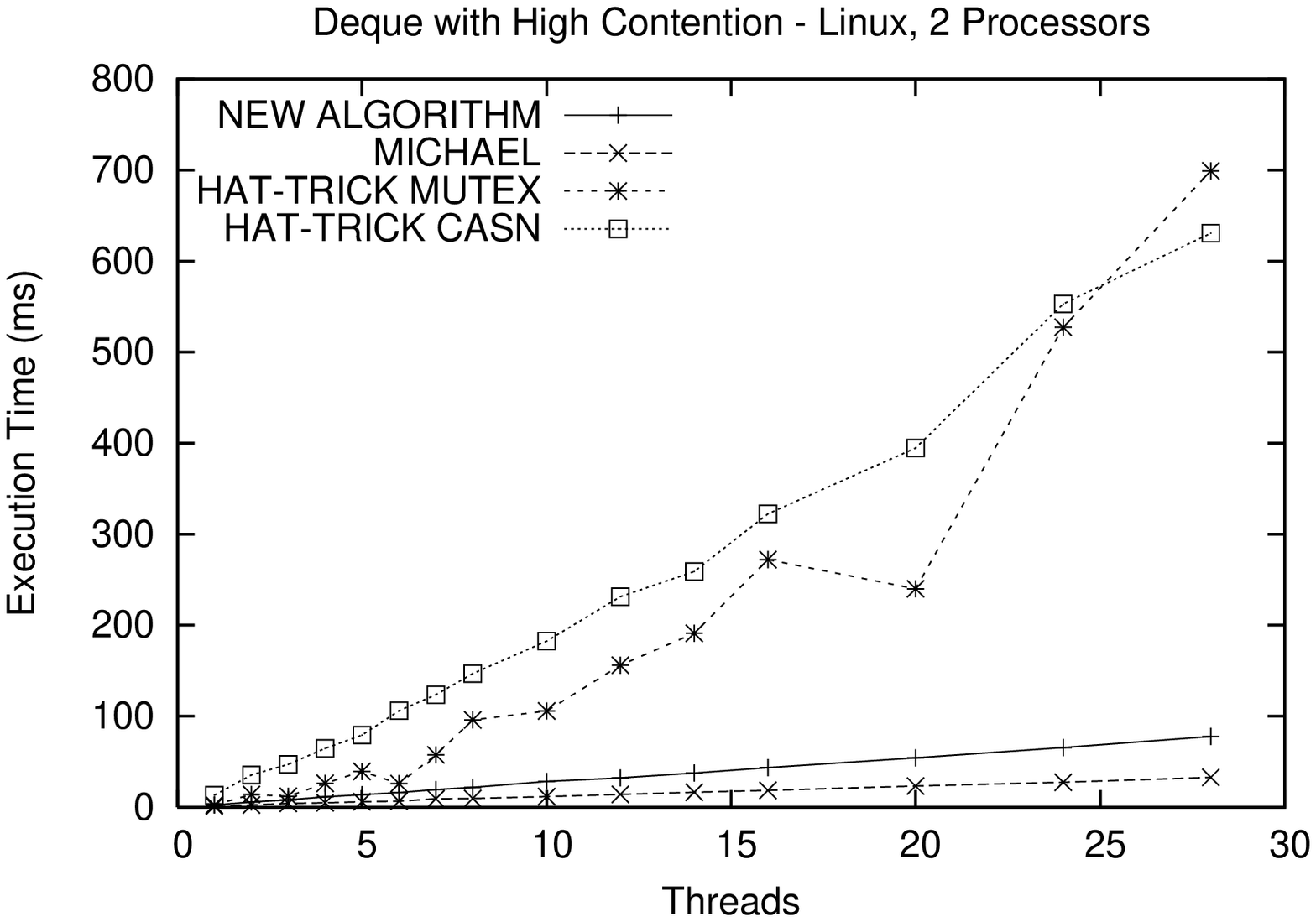, width=6.0cm, angle=270} &
\psfig{figure=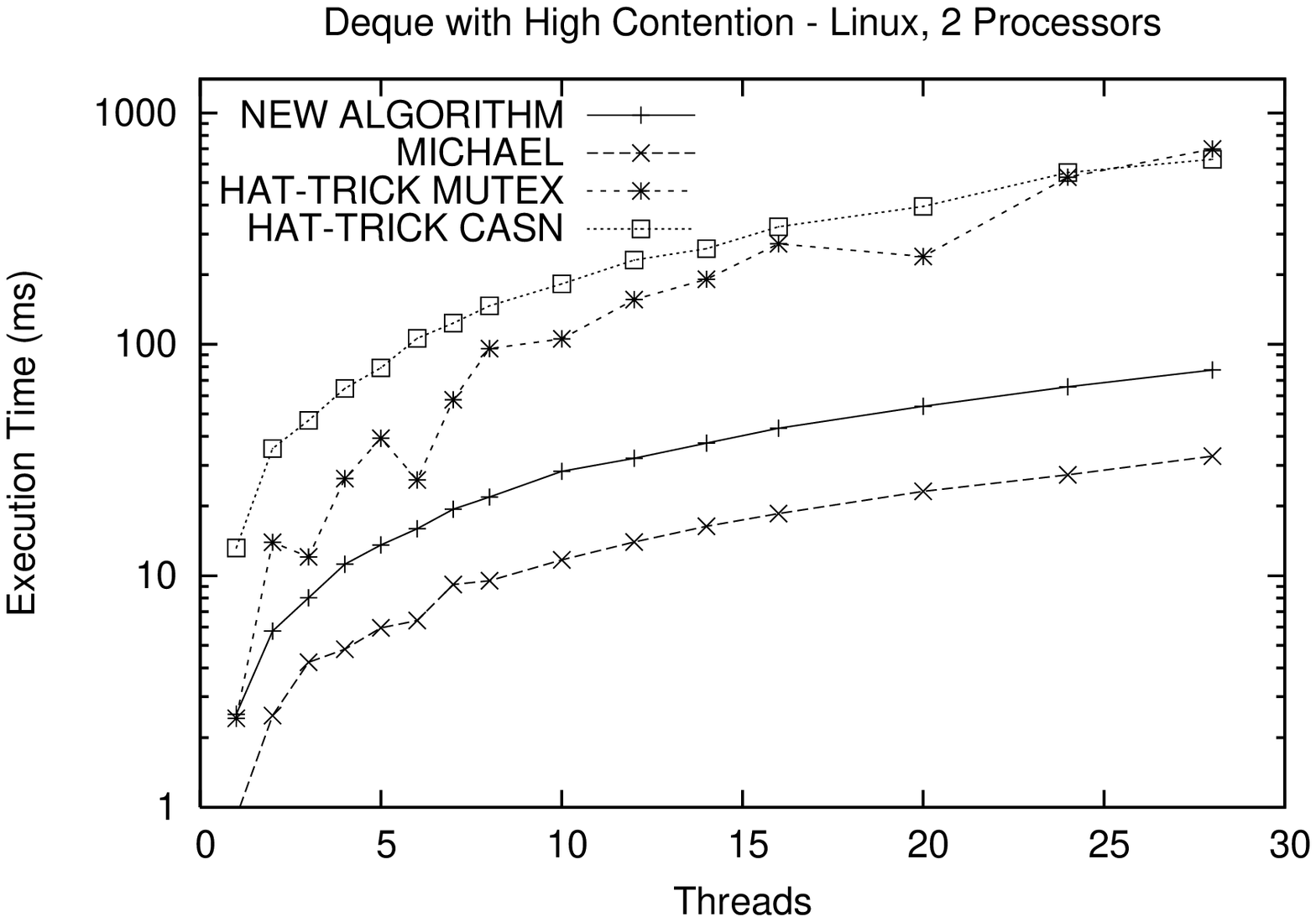, width=6.0cm, angle=270}\\
\end{tabular}
\end{center}
\caption{Experiment with deques and high contention. Logarithmic scales to the right.}
\label{fig:expDeque}
\end{figure*}

In our experiments, each concurrent thread performed 1000 randomly chosen sequential operations on a shared deque, with a distribution of 1/4 \textit{PushRight}, 1/4 \textit{PushLeft}, 1/4 \textit{PopRight} and 1/4 \textit{PopLeft} operations. Each experiment was repeated 50 times, and an average execution time for each experiment was estimated. Exactly the same sequential operations were performed for all different implementations compared. Besides our implementation, we also performed the same experiment with the lock-free implementation by Michael \cite{Mic03} and the implementation by Martin et al \cite{MarMS02}, two of the most efficient lock-free deques that have been proposed. The algorithm by Martin et al \cite{MarMS02} was implemented together with the corresponding memory management scheme by Detlefs et al \cite{DetMMS01}. However, as both \cite{MarMS02} and \cite{DetMMS01} use the atomic operation CAS2 which is not available in any modern system, the CAS2 operation was implemented in software using two different approaches. The first approach was to implement CAS2 using mutual exclusion (as proposed in \cite{MarMS02}), which should match the optimistic performance of an imaginary CAS2 implementation in hardware. The other approach was to implement CAS2 using one of the most efficient software implementations of CASN known that could meet the needs of \cite{MarMS02} and \cite{DetMMS01}, i.e. the implementation by Harris et al \cite{HarFP02}.

A clean-cache operation was performed just before each sub-experiment using a different implementation. All implementations are written in C and compiled with the highest optimization level. The atomic primitives are written in assembler.

The experiments were performed using different number of threads, varying from 1 to 28 with increasing steps. Three different platforms were used, with varying number of processors and level of shared memory distribution. To get a highly pre-emptive environment, we performed our experiments on a Compaq dual-processor Pentium II PC running Linux, and a Sun Ultra 80 system running Solaris 2.7 with 4 processors. In order to evaluate our algorithm with full concurrency we also used a SGI Origin 2000 system running Irix 6.5 with 29 250 MHz MIPS R10000 processors. The results from the experiments are shown in Figure \ref{fig:expDeque}. The average execution time is drawn as a function of the number of threads.
 
 Our results show that both the CAS-based algorithms outperforms the CAS2-based implementations for any number of threads. For the systems with low or medium concurrency and uniform memory architecture, \cite{Mic03} has the best performance. However, for the system with full concurrency and non-uniform memory architecture our algorithm performs significantly better than \cite{Mic03} from 2 threads and more, as a direct consequence of the disjoint-parallel accessible nature of our algorithm.

\section{Conclusions}\label{sect.conclusions}

We have presented the first lock-free algorithmic implementation of a concurrent deque that has all the following features: i) it is disjoint-parallel accessible with retained parallelism, ii) uses a fully described lock-free memory management scheme, and iii) uses atomic primitives which are available in modern computer systems, even when extended for dynamic maximum sizes.

We have performed experiments that compare the performance of our algorithm with two of the most efficient algorithms of lock-free deques known, using full implementations of those algorithms. The experiments show that our implementation performs significantly better on systems with high concurrency and non-uniform memory architecture.

We believe that our implementation is of highly practical interest for multi-processor applications. We are currently incorporating it into the NOBLE \cite{SunT02} library.


\bibliographystyle{IEEEtranS}
\bibliography{report}


\begin{appendix}

\section{The algorithm for dynamic maximum sizes}\label{sect.dynamic_algorithm}

This section shows the the full details of how to extend the algorithm for dynamic maximum sizes, following the guidelines earlier presented in Section \ref{sect.dynamicsize}.

As the link structure now can contain full pointer values, see Figure \ref{fig:dynamic_algorithm1} , the following functions are added for safe handling of the memory management:

\

\begin{small}

\textbf{function} READ\_NODE(address:\textbf{pointer to} Link) :\textbf{pointer to} Node

\textbf{function} READ\_DEL\_NODE(address:\textbf{pointer to} Link) :\textbf{pointer to} Node

\end{small}
\

The functions \textit{READ\_NODE} and \textit{READ\_DEL\_NODE} atomically de-references the given link and increases the reference counter for the corresponding node. In case the deletion mark of the link is set, the function \textit{READ\_NODE} returns NULL.

The remaining details of the extended algorithm are showed in Figures \ref{fig:dynamic_algorithm1} and \ref{fig:dynamic_algorithm2}.

\begin{figure*}
\begin{minipage}[t]{5.0cm}
\begin{small}
\begin{tabbing}
\ \ \ \ \ \ \ \ \ \ \ \=
          \ \ \ \ \= 
               \ \ \ \ \= 
                    \ \ \ \ \=
                         \ \ \ \ \=\\[-0.5cm]
\textbf{union} Link\\
\> \_: \textbf{word}\\
\> $\langle p,d \rangle$: $\langle$\textbf{pointer to} Node, \textbf{boolean}$\rangle$\\
\\
\textbf{structure} Node\\
\> value: \textbf{pointer to word}\\
\> prev: \textbf{union} Link\\
\> next: \textbf{union} Link\\
\\
// Global variables\\
head, tail: \textbf{pointer to} Node\\
// Local variables\\
node,prev,prev2,next,next2: \textbf{pointer to} Node\\
link1,lastlink: \textbf{union} Link\\
\\
\textbf{function} CreateNode(value: \textbf{pointer to word}):\textbf{pointer to} Node\\
C1  \> node:=MALLOC\_NODE();\\
C2  \> node.value:=value;\\
C3  \> \textbf{return} node;\\
\\
\textbf{procedure} ReleaseReferences(node: \textbf{pointer to} Node)\\
RR1 \> \> RELEASE\_NODE(node.prev.p);\\
RR2 \> \> RELEASE\_NODE(node.next.p);\\
\\
\textbf{procedure} PushLeft(value: \textbf{pointer to word})\\
L1 \> node:=CreateNode(value);\\
L2 \> prev:=COPY\_NODE(head);\\
L3 \> next:=READ\_NODE(\&prev.next);\\
L4 \> \textbf{while} \textbf{true} \textbf{do}\\
L5 \> \> \textbf{if} prev.next $\neq$ $\langle$next,\textbf{false}$\rangle$ \textbf{then}\\
L6 \> \> \> RELEASE\_NODE(next);\\
L7 \> \> \> next:=READ\_NODE(\&prev.next);\\
L8 \> \> \> \textbf{continue};\\
L9 \> \> node.prev:=$\langle$prev,\textbf{false}$\rangle$;\\
L10 \> \> node.next:=$\langle$next,\textbf{false}$\rangle$;\\
L11 \> \> \textbf{if} CAS(\&prev.next,$\langle$next,\textbf{false}$\rangle$,$\langle$node,\textbf{false}$\rangle$) \textbf{then}\\
L12 \> \> \> COPY\_NODE(node);\\
L13 \> \> \> \textbf{break};\\
L14 \> \> \textit{Back-Off}\\
L15 \> PushCommon(node,next);\\
\\
\textbf{procedure} PushRight(value: \textbf{pointer to word})\\
R1 \> node:=CreateNode(value);\\
R2 \> next:=COPY\_NODE(tail);\\
R3 \> prev:=READ\_NODE(\&next.prev);\\
R4 \> \textbf{while} \textbf{true} \textbf{do}\\
R5 \> \> \textbf{if} prev.next $\neq$ $\langle$next,\textbf{false}$\rangle$ \textbf{then}\\
R6 \> \> \> prev:=HelpInsert(prev,next);\\
R7 \> \> \> \textbf{continue};\\
R8 \> \> node.prev:=$\langle$prev,\textbf{false}$\rangle$;\\
R9 \> \> node.next:=$\langle$next,\textbf{false}$\rangle$;\\
R10 \> \> \textbf{if} CAS(\&prev.next,$\langle$next,\textbf{false}$\rangle$,$\langle$node,\textbf{false}$\rangle$) \textbf{then}\\
R11 \> \> \> COPY\_NODE(node);\\
R12 \> \> \> \textbf{break};\\
R13 \> \> \textit{Back-Off}\\
R14 \> PushCommon(node,next);
\end{tabbing}
\end{small}
\end{minipage}
\hfill
\begin{minipage}[t]{5.0cm}
\begin{small}
\begin{tabbing}
\ \ \ \ \ \ \ \ \ \ \ \=
          \ \ \ \ \= 
               \ \ \ \ \= 
                    \ \ \ \ \=
                         \ \ \ \ \=\\[-0.5cm]
\textbf{procedure} PushCommon(node, next: \textbf{pointer to} Node)\\
P1 \> \textbf{while} \textbf{true} \textbf{do}\\
P2 \> \> link1:=next.prev;\\
P3 \> \> \textbf{if} link1.d $=$ \textbf{true} \textbf{or} node.next $\neq$ $\langle$next,\textbf{false}$\rangle$ \textbf{then}\\
P4 \> \> \> \textbf{break};\\
P5 \> \> \textbf{if} CAS(\&next.prev,link1,$\langle$node,\textbf{false}$\rangle$) \textbf{then}\\
P6 \> \> \> COPY\_NODE(node);\\
P7 \> \> \> RELEASE\_NODE(link1.p);\\
P8 \> \> \> \textbf{if} node.prev.d = \textbf{true} \textbf{then}\\
P9 \> \> \> \> prev2:=COPY\_NODE(node);\\
P10 \> \> \> \> prev2:=HelpInsert(prev2,next);\\
P11 \> \> \> \> RELEASE\_NODE(prev2);\\
P12 \> \> \> \textbf{break};\\
P13 \> \> \textit{Back-Off}\\
P14 \> RELEASE\_NODE(next);\\
P15 \> RELEASE\_NODE(node);\\
\\
\textbf{function} PopLeft(): \textbf{pointer to word}\\
PL1 \> prev:=COPY\_NODE(head);\\
PL2 \> \textbf{while} \textbf{true} \textbf{do}\\
PL3 \> \> node:=READ\_NODE(\&prev.next);\\
PL4 \> \> \textbf{if} node $=$ tail \textbf{then} \\
PL5 \> \> \> RELEASE\_NODE(node);\\
PL6 \> \> \> RELEASE\_NODE(prev);\\
PL7 \> \> \> \textbf{return} $\bot$;\\
PL8 \> \> link1:=node.next;\\
PL9 \> \> \textbf{if} link1.d $=$ \textbf{true} \textbf{then}\\
PL10 \> \> \> DeleteNext(node);\\
PL11 \> \> \> RELEASE\_NODE(node);\\
PL12 \> \> \> \textbf{continue};\\
PL13 \> \> \textbf{if} CAS(\&node.next,link1,$\langle$link1.p,\textbf{true}$\rangle$) \textbf{then}\\
PL14 \> \> \> DeleteNext(node);\\
PL15 \> \> \> next:=READ\_DEL\_NODE(\&node.next);\\
PL16 \> \> \> prev:=HelpInsert(prev,next);\\
PL17 \> \> \> RELEASE\_NODE(prev);\\
PL18 \> \> \> RELEASE\_NODE(next);\\
PL19 \> \> \> value:=node.value;\\
PL20 \> \> \> \textbf{break};\\
PL21 \> \> RELEASE\_NODE(node);\\
PL22 \> \> \textit{Back-Off}\\
PL23 \> RemoveCrossReference(node);\\
PL24 \> RELEASE\_NODE(node);\\
PL25 \> \textbf{return} value;\\
\\
\textbf{function} PopRight(): \textbf{pointer to word}\\
PR1 \> next:=COPY\_NODE(tail);\\
PR2 \> node:=READ\_NODE(\&next.prev);\\
PR3 \> \textbf{while} \textbf{true} \textbf{do}\\
PR4 \> \> \textbf{if} node.next $\neq$ $\langle$next,\textbf{false}$\rangle$ \textbf{then}\\
PR5 \> \> \> node:=HelpInsert(node,next);\\
PR6 \> \> \> \textbf{continue};\\

PR7 \> \> \textbf{if} node $=$ head \textbf{then} \\
PR8 \> \> \> RELEASE\_NODE(node);\\
PR9 \> \> \> RELEASE\_NODE(next);\\
PR10 \> \> \> \textbf{return} $\bot$;\\
\end{tabbing}
\end{small}
\end{minipage}
\caption{The algorithm for dynamic maximum sizes, part 1(2).}
\label{fig:dynamic_algorithm1}
\end{figure*}

\begin{figure*}
\begin{minipage}[t]{5.0cm}
\begin{small}
\begin{tabbing}
\ \ \ \ \ \ \ \ \ \ \ \=
          \ \ \ \ \= 
               \ \ \ \ \= 
                    \ \ \ \ \=
                         \ \ \ \ \=\\[-0.5cm]
PR11 \> \> \textbf{if} CAS(\&node.next,$\langle$next,\textbf{false}$\rangle$,$\langle$next,\textbf{true}$\rangle$) \textbf{then}\\
PR12 \> \> \> DeleteNext(node);\\
PR13 \> \> \> prev:=READ\_DEL\_NODE(\&node.prev);\\
PR14 \> \> \> prev:=HelpInsert(prev,next);\\
PR15 \> \> \> RELEASE\_NODE(prev);\\
PR16 \> \> \> RELEASE\_NODE(next);\\
PR17 \> \> \> value:=node.value;\\
PR18 \> \> \> \textbf{break};\\
PR19 \> \> \textit{Back-Off}\\
PR20 \> RemoveCrossReference(node);\\
PR21 \> RELEASE\_NODE(node);\\
PR22 \> \textbf{return} value;\\
\\
\textbf{procedure} DeleteNext(node: \textbf{pointer to} Node)\\
DN1 \> \textbf{while} \textbf{true} \textbf{do}\\
DN2 \> \> link1:=node.prev;\\
DN3 \> \> \textbf{if} link1.d $=$ \textbf{true} \textbf{or}\\ 
DN4 \> \> ~ CAS(\&node.prev,link1,$\langle$link1.p,\textbf{true}$\rangle$) \textbf{then} \textbf{break};\\
DN5 \> lastlink.d:=\textbf{true};\\
DN6 \> prev:=READ\_DEL\_NODE(\&node.prev);\\
DN7 \> next:=READ\_DEL\_NODE(\&node.next);\\
DN8 \> \textbf{while} \textbf{true} \textbf{do}\\
DN9 \> \> \textbf{if} prev $=$ next \textbf{then} \textbf{break};\\
DN10 \> \> \textbf{if} next.next.d = \textbf{true} \textbf{then}\\
DN11 \> \> \> next2:=READ\_DEL\_NODE(\&next.next);\\
DN12 \> \> \> RELEASE\_NODE(next);\\
DN13 \> \> \> next:=next2;\\
DN14 \> \> \> \textbf{continue};\\
DN15 \> \> prev2:=READ\_NODE(\&prev.next);\\
DN16 \> \> \textbf{if} prev2 $=$ NULL \textbf{then}\\
DN17 \> \> \> \textbf{if} lastlink.d = \textbf{false} \textbf{then}\\
DN18 \> \> \> \> DeleteNext(prev);\\
DN19 \> \> \> \> lastlink.d:=\textbf{true};\\
DN20 \> \> \> prev2:=READ\_DEL\_NODE(\&prev.prev);\\
DN21 \> \> \> RELEASE\_NODE(prev);\\
DN22 \> \> \> prev:=prev2;\\
DN23 \> \> \> \textbf{continue};\\
DN24 \> \> \textbf{if} prev2 $\neq$ node \textbf{then}\\
DN25 \> \> \> lastlink.d:=\textbf{false};\\
DN26 \> \> \> RELEASE\_NODE(prev);\\
DN27 \> \> \> prev:=prev2;\\
DN28 \> \> \> \textbf{continue};\\
DN29 \> \> RELEASE\_NODE(prev2);\\
DN30 \> \> \textbf{if} CAS(\&prev.next,$\langle$node,\textbf{false}$\rangle$,$\langle$next,\textbf{false}$\rangle$) \textbf{then}\\
DN31 \> \> \> COPY\_NODE(next);\\
DN32 \> \> \> RELEASE\_NODE(node);\\
DN33 \> \> \> \textbf{break};\\
DN34 \> \> \textit{Back-Off}\\
DN35 \> RELEASE\_NODE(prev);\\
DN36 \> RELEASE\_NODE(next);
\end{tabbing}
\end{small}
\end{minipage}
\hfill
\begin{minipage}[t]{5.0cm}
\begin{small}
\begin{tabbing}
\ \ \ \ \ \ \ \ \ \ \ \=
          \ \ \ \ \= 
               \ \ \ \ \= 
                    \ \ \ \ \=
                         \ \ \ \ \=\\[-0.5cm]
\textbf{function} HelpInsert(prev, node: \textbf{pointer to} Node)\\
~ :\textbf{pointer to} Node\\
HI1\> lastlink.d:=\textbf{true};\\
HI2\> \textbf{while} \textbf{true} \textbf{do}\\
HI3\> \> prev2:=READ\_NODE(\&prev.next);\\
HI4\> \> \textbf{if} prev2 $=$ NULL \textbf{then}\\
HI5\> \> \> \textbf{if} lastlink.d = \textbf{false} \textbf{then}\\
HI6\> \> \> \> DeleteNext(prev);\\
HI7\> \> \> \> lastlink.d:=\textbf{true};\\
HI8\> \> \> prev2:=READ\_DEL\_NODE(\&prev.prev);\\
HI9\> \> \> RELEASE\_NODE(prev);\\
HI10\> \> \> prev:=prev2;\\
HI11\> \> \> \textbf{continue};\\
HI12\> \> link1:=node.prev;\\
HI13\> \> \textbf{if} link1.d $=$ \textbf{true} \textbf{then}\\
HI14\> \> \> RELEASE\_NODE(prev2);\\
HI15\> \> \> \textbf{break};\\
HI16\> \> \textbf{if} prev2 $\neq$ node \textbf{then}\\
HI17\> \> \> lastlink.d:=\textbf{false};\\
HI18\> \> \> RELEASE\_NODE(prev);\\
HI19\> \> \> prev:=prev2;\\
HI20\> \> \> \textbf{continue};\\
HI21\> \> RELEASE\_NODE(prev2);\\
HI22\> \> \textbf{if} CAS(\&node.prev,link1,$\langle$prev,\textbf{false}$\rangle$) \textbf{then}\\
HI23\> \> \> COPY\_NODE(prev);\\
HI24\> \> \> RELEASE\_NODE(link1.p);\\
HI25\> \> \> \textbf{if} prev.prev.d $=$ \textbf{true} \textbf{then} \textbf{continue};\\
HI26\> \> \> \textbf{break};\\
HI27\> \> \textit{Back-Off}\\
HI28\> \textbf{return} prev;\\
\\
\textbf{procedure} RemoveCrossReference(node: \textbf{pointer to} Node)\\
RC1\> \textbf{while} \textbf{true} \textbf{do}\\
RC2\> \> prev:=node.prev.p;\\
RC3\> \> \textbf{if} prev.next.d = \textbf{true} \textbf{then}\\
RC4\> \> \> prev2:=READ\_DEL\_NODE(\&prev.prev);\\
RC5\> \> \> node.prev:=$\langle$prev2,\textbf{true}$\rangle$;\\
RC6\> \> \> RELEASE\_NODE(prev);\\
RC7\> \> \> \textbf{continue};\\
RC8\> \> next:=node.next.p;\\
RC9\> \> \textbf{if} next.next.d = \textbf{true} \textbf{then}\\
RC10\> \> \> next2:=READ\_DEL\_NODE(\&next.next);\\
RC11\> \> \> node.next:=$\langle$next2,\textbf{true}$\rangle$;\\
RC12\> \> \> RELEASE\_NODE(next);\\
RC13\> \> \> \textbf{continue};\\
RC14\> \> \textbf{break};\\
\end{tabbing}
\end{small}
\end{minipage}
\caption{The algorithm for dynamic maximum sizes, part 2(2).}
\label{fig:dynamic_algorithm2}
\end{figure*}

\end{appendix}


\end{document}